\documentstyle[preprint,aps]{revtex}
\tightenlines
\input  epsf.tex

\begin{document}
\title{M\O LLER ENERGY FOR THE KERR-NEWMAN METRIC}
\author{S.~S.~Xulu\thanks{%
E-mail: ssxulu@pan.uzulu.ac.za}}
\address{Department of Applied Mathematics, University of Zululand,\\
Private Bag X1001,3886 Kwa-Dlangezwa, South Africa}
\maketitle

\begin{abstract}
       The energy distribution in the Kerr-Newman space-time is computed 
using the M\o ller energy-momentum complex. This  agrees with the Komar mass 
for this space-time obtained by Cohen and de Felice. These results support 
the Cooperstock hypothesis.
\end{abstract}


\section{Introduction}

The notion of energy and momentum localization has been associated with much 
debate since  the advent of the  general theory of relativity (see 
\cite{debate} and references therein). About  a decade  ago a renowned 
general relativist, H. Bondi\cite{Bondi}, argued that a nonlocalizable form 
of energy is not 
allowed  in relativity and therefore its location can in principle be found.
In a flat space-time the concept of energy localization is not controversial. 
In this case the energy-momentum tensor $T_{i}^{\ k}$ satisfies the divergence
relation $T_{i\ ,k}^{\ k}=0$. The presence of gravitation however  
necessitates the replacement of an ordinary derivative by a covariant one, 
and this leads to the covariant conservation laws $T_{i\ ;\ k}^{\ k}=0$. In 
a curved space-time the energy-momentum tensor of matter plus all 
non-gravitational fields no longer  satisfies $T_{i\text{ },\text{ }k}^{\ k}=0$; 
the contribution from the gravitational field is now required to construct an 
energy-momentum expression which satisfies a divergence relation like one has 
in a flat space-time. Attempts aimed at obtaining a meaningful expression for
energy, momentum, and angular momentum  for a general relativistic  system 
resulted in many different definitions (see references in \cite{ACV,Vir99}).
Einstein's energy-momentum complex,
used for calculating the energy  in a general relativistic
system, was followed by many prescriptions: e.g. Landau and Lifshitz,
Papapetrou and Weinberg. These energy-momentum complexes restrict one
to make calculations in ``Cartesian coordinates''. 
This shortcoming of having to single out a particular coordinate system 
prompted M\o ller\cite{Moller58}
to construct an expression which enables one to evaluate energy in any
coordinate system. M\o ller claimed that his\ expression gives the same
values for the total energy and momentum as the Einstein's energy-momentum
complex for a closed system. However, 
M\o ller's energy-momentum complex was subjected to some criticism
(see in  \cite{MolCrit}). Further Komar\cite{Komar} formulated a new
definition of energy in a curved space-time. This prescription, though
not restricted to  the use of ``Cartesian coordinates'', is not applicable
to non-static space-times.

 A large number of definitions of quasi-local mass  have been 
proposed (see  \cite{Penrose,BrownYork,Hayw} and references therein). The
uses of quasi-local masses to obtain energy in a curved space-time are not
limited to a particular coordinates system whereas many energy-momentum
complexes are restricted to the use of ``Cartesian coordinates.''  
Penrose \cite{Penrose} pointed out that  quasi-local masses  are conceptually 
very important.
However, inadequacies of these quasi-local masses (these different definitions
 do not give agreed
results for the Reissner-Nordstr\"{o}m and Kerr metrics and  that the Penrose 
definition  could not succeed to deal with the Kerr metric)  have been
discussed in \cite{BergBernTod,Vir99}. Contrary to this, Virbhadra, his
collaborators and others\cite{VirKN,CoopRi,VirOthFlat} considered many asymptotically
flat space-times and showed  that several energy-momentum complexes give the
same and acceptable results for a given space-time. 
Further Rosen and Virbhadra, Virbhadra, and some others  carried out
calculations on a few asymptotically non-flat space-times using different 
energy-momentum complexes and got encouraging results\cite{VirOthNonFlat}. 
Aguirregabiria {\it et al.} \cite{ACV} proved that several energy-momentum complexes give the same
result for any Kerr-Schild class metric. Virbhadra\cite{Vir99}  also
showed that for a general nonstatic, spherically symmetric space-time of the
Kerr-Schild class the Penrose quasilocal mass definition as well as several
energy-momentum complexes yield the same results. Recently, Chang {\it et al.
}\cite{Changetal}  showed that every energy-momentum complex can be
associated with a particular Hamiltonian boundary term. Therefore the
energy-momentum complexes may also be considered as quasi-local.

The energy distribution in the Kerr-Newman space-time was earlier evaluated by
Cohen and de Felice \cite{CodeFe} using Komar's prescription. Virbhadra\cite
{VirKN} showed that, up to the third order of the rotation parameter, the
definitions of Einstein and Landau-Lifshitz give the same and reasonable
energy distribution in the Kerr-Newman (KN) field when calculations are
carried out in Kerr-Schild Cartesian coordinates.
 Cooperstock and Richardson 
\cite{CoopRi} extended the Virbhadra energy calculations up to the seventh
order of the rotation parameter and found that these definitions give the
same energy distribution for the KN metric.
 Aguirregabiria {\it et al.} \cite{ACV} performed exact  
computations  for the energy distribution in  KN space-time in Kerr-Schild
Cartesian coordinates. They   showed that the energy distribution in the
prescriptions of Einstein, Landau-Lifshitz, Papapetrou, and Weinberg (ELLPW)
gave the same result. 

In a  recent paper  Lessner\cite{Lessner}\ in his
analysis of  M\o ller's energy-momentum expression concludes that it  is
a powerful representation  of energy and momentum in general relativity. Therefore,
it is  interesting and important to obtain  energy distribution  using 
M\o ller's prescription. In this paper we evaluate the energy distribution for
the KN space-time in M\o ller's\cite{Moller58} prescription and compare the
result with those already obtained using Komar's mass as well as the ELLPW 
energy-momentum complexes. We are also interested to check whether or not
the {\em Cooperstock hypothesis}\cite{CoopHypoth} (which essentially states that the energy and 
momentum in a curved space-time  are confined to the regions of non-vanishing
energy-momentum tensor $T_i^{\ k}$ of the matter and all non-gravitational 
fields) holds good for this case.
 We use the convention that Latin indices take 
values from $0$ to $ 3$ and Greek indices values from $1$ to $3$, and take 
$G=1$ and $c=1$ units. 

\section{The Kerr-Newman metric}

The stationary  axially symmetric and asymptotically flat Kerr-Newman 
solution is 
the most general black hole solution to the Einstein-Maxwell equations.
This  describes the exterior gravitational and electromagentic field  of a
charged rotating object. The  Kerr-Newman metric  in Boyer-Lindquist 
coordinates $(t,\rho ,\theta ,\phi )$ is expressed by the line element

\begin{equation}
ds^{2}=\frac{\Delta }{r_{0}^{2}}[dt-a\sin ^{2}\theta d\phi ]^{2}-\frac{
\sin ^{2}\theta }{r_{0}^{2}}[\left( \rho ^{2}+a^{2}\right) d\phi
-a dt]^{2}-\frac{r_{0}^{2}}{\Delta }d\rho ^{2}-r_{0}^{2}d\theta ^{2}.
\label{KNMetricBL}
\end{equation}
where $\Delta := \rho^2- 2M \rho + e^2+a^2$  and $r_{0}^{2} := 
\rho^{2}+a^{2}\cos^{2}\theta $. $M$, $e$ and $a$ are  respectively 
{\em  mass}, {\em electric charge} and {\em  rotation} parameters.
This space-time has $\{\rho = constant\}$ null hypersurfaces for
$g^{\rho \rho}=0$, which are given by
\begin{equation}
\rho _{\pm}=M \pm \sqrt{M^{2}-e^{2}-a^{2}}  \text{\ .} \label{nullhyp}
\end{equation}
There is a ring curvature singularity $r_0 =0$ in the KN space-time.
This  space-time has an event horizon at $\rho = \rho_+$; 
 it describes a black hole if and only if $M^{2}\geq e^{2}+a^{2}$. 
The coordinates are singular at $\rho = \rho_{\pm}$. Therefore, $t$
is replaced with a null coordinate $v$ and $\phi$ with 
${\varphi }$ by the following transformation: 
\begin{eqnarray}
dt     &=&  dv - \frac{\rho ^{2}+a^{2}}{\Delta }d\rho ,  \nonumber \\
d\phi  &=&  d\varphi - \frac{a}{\Delta }d\rho 
\end{eqnarray}
and thus the  KN metric   is expressed in advanced 
Eddington-Finkelstein coordinates $(v, \rho, \theta, \varphi)$ 	as
\begin{eqnarray}
ds^{2} 
&=&
    \left( 1-\frac{2M\rho}{r_{0}^{2}}+\frac{e^{2}}{r_{0}^{2}}\right)dv^{2}
    -2dv\ d\rho 
    +\frac{2a\sin^2 \theta} {r_{0}^{2}}\left( 2M\rho-e^{2}\right) dv\ d\varphi
   -r_{0}^{2}d\theta^{2}  \nonumber \\
&&
    + 2a \sin^{2} \theta d\rho  d\varphi 
  -\left[
    \left(\rho^2+a^2\right)\sin^2\theta
    + \frac{2M \rho-e^2}{{r_o}^2}  a^2 \sin^4\theta
   \right] d\varphi^2 \text{.}
\end{eqnarray}

Transforming the above to  Kerr-Schild Cartesian coordinates $(T,x,y,z)$
according to
\begin{eqnarray}
T &=& v - \rho \nonumber\\
x &=& \sin\theta \left(\rho \cos\varphi + a \sin\varphi\right) \nonumber\\
y &=& \sin\theta \left(\rho \sin\varphi - a \cos\varphi\right) \nonumber\\
z &=& \rho \cos\theta
\end{eqnarray}
one has the line element
\begin{eqnarray}
ds^2 &=& dT^2 - dx^2 -dy^2 -dz^2 
    - \frac{\left(2m\rho-e^2\right)\rho^2}{\rho^4+a^2 z^2}  \times \nonumber\\
     && \left(
              dT 
             + \frac{\rho}{a^2+\rho^2} \left(x dx + y dy\right)
             + \frac{a}{a^2+\rho^2} \left(y dx - x dy\right)
             +\frac{z}{\rho} dz 
      \right)^2 \text{.}
\end{eqnarray}
\section{Energy distribution in Kerr-Newman metric.}

In this Section we first give the energy distribution in the KN
space-time obtained by some authors  and then using  the M\o ller 
energy-momentum complex we obtain the energy distribution for the same 
space-time.

The energy distribution in Komar's prescription obtained by
Cohen and de Felice\cite{CodeFe}, using the KN metric in Boyer-Lindquist
coordinates, is given by 
\begin{equation}
E_{\rm  K} = M-\frac{e^{2}}{2\rho }\left[ 1+\frac{(a^{2}+\rho^{2})}{a\rho }
\arctan \left( \frac{a}{\rho }\right) \right] \text{.}  \label{EKomar}
\end{equation}
(The subscript K on the left hand side of the equation refers to Komar.)
Aguirregabiria {\it  et al.}\cite{ACV}  studied the energy-momentum complexes
of Einstein, Landau-Lifshitz, Papapetrou and Weinberg for the KN metric.
They showed that these definitions give the same results for the energy
and energy current densities. They used the KN metric in Kerr-Schild 
Cartesian coordinates. They found that these definitions give the
same result for the energy distributon  for the KN metric, which is
expressed as
\begin{equation}
E_{\rm ELLPW} = M-\frac{e^{2}}{4\rho }\left[ 1+\frac{(a^{2}+\rho^{2})}{a\rho }%
\arctan \left( \frac{a}{\rho }\right) \right] \text{.}  \label{EELLPW}
\end{equation}
(The subscript ELLPW on the left hand side of the above equation refers to 
the Einstein, Landau-Lifshitz, Papapetrou and Weinberg prescriptions.)
It is obvious that the Komar definition gives a different result for the
Kerr-Newman metric as compared to those  obtained using energy-momentum 
complexes of ELLPW. However, for the Kerr metric ($e=0$)  all these definitions
yield the same results. These results obviously support the Cooperstock
hypothesis.
The M\o ller  energy-momentum complex ${\Im }_{i}^{\ k}$\ is given
by \cite{Moller58}
\begin{equation}
{\Im }_{i}^{\ k}=\frac{1}{8\pi }{\chi}_{i\ \ ,l}^{\ kl} \quad \text{,} 
 \label{MollerEMC}
\end{equation}
satisfying the local conservation laws: 
\begin{equation}
\frac{\partial {\Im }_{i}^{\ k}}{\partial x^{k}}= 0 \text{ ,}
\label{ConsvLaws}
\end{equation}
where the antisymmetric superpotential ${\chi}_{i}^{\ kl}$ is  
\begin{equation}
{\chi}_{i}^{\ kl} = - {\chi}_{i}^{\ lk} = \sqrt{-g} \left[
g_{in,m}-g_{im,n}\right]  g^{km}g^{nl}\text{ \ .}  
\label{Chi}
\end{equation}
 The energy and momentum components  are given by
\begin{equation}
P_{i}=\int \int \int {\Im }_{i}^{\ 0} dx^{1} dx^{2} dx^{3} \text{\ ,}
\end{equation}
where $P_{0}$ is the energy  while $P_{\alpha }$ stand for
the momentum  components. Using  Gauss's theorem the energy $E$ for
a stationary metric is thus given by  
\begin{equation}
E = \frac{1}{8\pi } \int \int {\chi}_{0}^{\ 0\beta }\ \mu _{\beta }\ dS
\label{energy}
\end{equation}
where $\mu _{\beta }$ is the outward unit normal vector over an
infinitesimal surface element $dS$.

The  only  required  component of $\chi_i^{\ kl}$ is
\begin{equation}
{\chi}_{0}^{\ 01}=\frac{-2(\rho ^{2}+a^{2})\sin \theta }{(\rho
^{2}+a^{2}\cos ^{2}\theta )^{2}}\left( M a^{2}\cos ^{2}\theta -M\rho^{2}
 +e^{2}\rho \right)  \text{.}
\end{equation}
Using the above expression in equation  $(\ref{energy})$ the energy $E$
 inside a surface with  $\{\rho=constant\}$\  is then given by
\begin{equation}
E_{\rm M\o l} = M-\frac{e^{2}}{2\rho }\left[ 1+\frac{(a^{2}+\rho ^{2})}{a\rho }
    \arctan \left( \frac{a}{\rho }\right) \right] \text{.}  
\label{EMol}
\end{equation}
(The subscript M\o l  on the left hand side of this equation refers to M\o ller's
prescription.)  Our result, using M\o ller's complex, agrees with the energy 
distribution obtained by Cohen and de Felice\cite{CodeFe} in Komar's 
prescription. The  second term of the energy distribution differs by a 
factor of two from that  computed by Aguirregabiria {\it et al.} using  
ELLPW complexes. In both cases the energy is shared by both the interior 
and exterior of the KN black hole. It is clear that the definitions of 
ELLPW and Komar, M\o ller's definition also upholds the Cooperstock
hypothesis for the KN metric. The total energy ($ \rho \rightarrow \infty$
in all these energy expressions) give the same result $M$. 
\begin{figure*}
\epsfxsize 12cm
\epsfbox{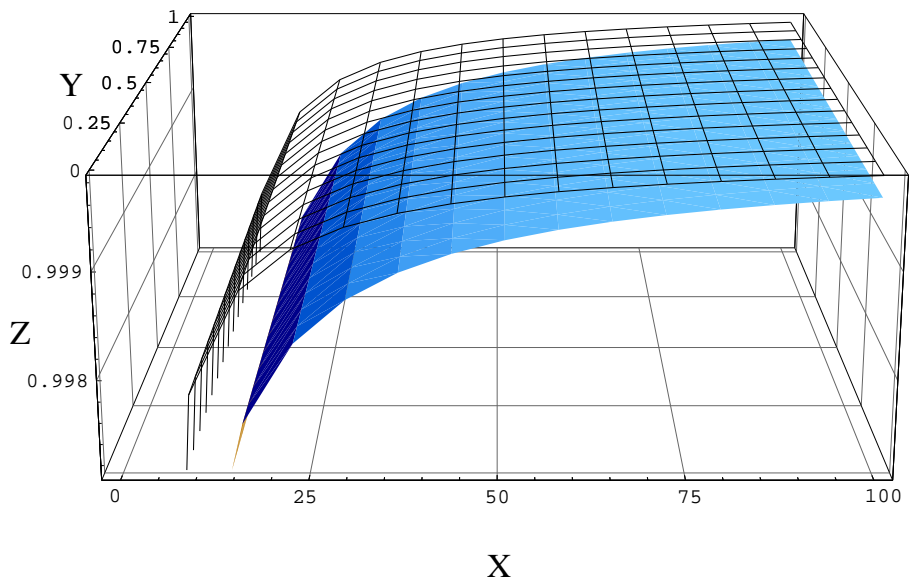}
\caption{
${\cal E}_{\rm ELLPW}$ and  ${\cal E}_{\rm KM}$ on Z-axis are 
    plotted against  ${\cal R}$ on X-axis and ${\cal S}$ on Y-axis for 
   ${\cal Q} = 0.2$. The upper (grid like) and lower surfaces 
 are for ${\cal E}_{\rm ELLPW}$ and  ${\cal E}_{\rm KM}$ respectively.
}
\label{fig1}
\end{figure*}
Now defining 
\begin{eqnarray}
&&{\cal E}_{\rm ELLPW} :=  \frac{E_{\rm ELLPW}}{M},  \quad
E_{\rm KM} := E_{\rm K} = E_{\rm M\o l},  \quad
{\cal E}_{\rm KM}  := \frac{E_{\rm KM}}{M} ,  \nonumber\\
&&{\cal S}  :=   \frac{a}{M},  \quad
{\cal Q}  :=    \frac{e}{M},  \quad
{\cal R}  :=   \frac{\rho}{M}  
\end{eqnarray}
the equations  $(\ref{EKomar})$,  $(\ref{EELLPW})$ and $(\ref{EMol})$
may be expressed as  
\begin{equation}
{\cal E}_{\rm ELLPW} =
    1-\frac{{\cal Q}^{2}}{4{\cal R}}\left[ 1+\left( \frac{{\cal S}}{{\cal R}}
   +\frac{{\cal R}}{{\cal S}}\right)
    \arctan \left( \frac{{\cal S}}{{\cal R}}\right) \right]  
\label{CalEELLPW}
\end{equation}
and 
\begin{equation}
{\cal E}_{\rm KM} =
    1-\frac{{\cal Q}^2}{2{\cal R}}\left[ 1+\left( \frac{{\cal S}}{{\cal R}}
    +\frac{{\cal R}}{{\cal S}}\right)
    \arctan \left( \frac{{\cal S}}{{\cal R}}\right) \right]  \text{\ .}
\label{CalEKM}
\end{equation}
The ring curvature singularity  in the KN metric is covered by the
event horizon for $\left( {\cal Q}^2+{\cal S}^2\right) \leq 1$ and is
naked for $\left( {\cal Q}^2+{\cal S}^2\right) >  1$. In Fig. 1 we plot
${\cal E}_{\rm ELLPW}$ and  ${\cal E}_{\rm KM}$ against ${\cal R}$
and ${\cal S}$ for ${\cal Q} = 0.2$. As  the value of  ${\cal R}$ increases 
the two  surfaces shown in the figure come closer.

\section{Conclusions}
The use of  the Komar definition is not restricted to ``Cartesian coordinates''
(like one has for the ELLPW complexes); however, it is applicable only to 
the stationary space-times. The M\o ller energy-momentum complex is
neither restricted to the use of particular coordinates nor to the stationary
space-times. Lessner\cite{Lessner} pointed out that the M\o ller defintion is
a powerful representation  of energy and momentum in general relativity. However,
one finds  that for the Reissner-Nordstr\"{o}m  metric 
$E_{\rm ELLPW} = M - e^2/(2 \rho)$ (the Penrose definition also gives the same
result which agrees with linear theory\cite{Tod} whereas  $E_{\rm KM} = M - e^2/\rho$.
Therefore, one prefers the results obtained by  using the definitions of
ETLLPW. It is   worth investigating
the Cooperstock hypothesis for  non-static space-times with M\o ller's
energy-momentum complex.

\acknowledgments
I am grateful to my supervisor K. S. Virbhadra for his guidance and NRF 
for financial support.


\begin{references}
\bibitem{debate} 
     C. W. Misner, K. S. Thorne and J. A. Wheeler, {\em Gravitation} 
          (W. H. Freeman and Co., NY, 1973) p.603; 
     F. I. Cooperstock and R. S. Sarracino, {\em J. Phys. }{\bf A11}, 877 
          (1978);
     S. Chandrasekhar and V. Ferrari, {\em Proceedings of Royal Society of 
        London} {\bf A435},  645 (1991).

\bibitem{Bondi}  H. Bondi, {\em Proc. Roy. Soc. London} {\bf A427}, 249
               (1990).
\bibitem{ACV}  J. M. Aguirregabiria, A. Chamorro, and K. S. Virbhadra, {\em 
             Gen. Relativ. Gravit.}. {\bf 28}, 1393 (1996).

\bibitem{Vir99}  K. S. Virbhadra, {\em Phys. Rev.} {\bf D60}, 104041 (1999).

\bibitem{Moller58} C. M\o ller, {\em Annals of Physics} (NY) {\bf 4}, 347
                  (1958).
\bibitem{MolCrit}  C. M\o ller, {\em Annals of Physics} (NY) {\bf 12},
                  118 (1961); 
          D. Kovacs, {\em Gen. Relativ. Gravit}. {\bf 17}, 927 (1985); 
          J. Novotny, {\em Gen. Relativ. Gravit}. {\bf 19}, 1043 (1987).
\bibitem{Komar}  A. Komar, {\em Phy. Rev. }{\bf 113,} 934 (1959).
\bibitem{Penrose}  R. Penrose, {\em Proc. Roy. Soc. London} {\bf A381}, 53
                 (1982).
\bibitem{BrownYork} J. D. Brown and J. W. York, Jr., {\em Phys. Rev.} {\bf D47},
                1407 (1993).
\bibitem{Hayw}  S. A. Hayward, {\em Phys. Rev.} {\bf D49}, 831 (1994).
\bibitem{BergBernTod} 
    G. Bergqvist, {\em Class. Quantum Gravit.} {\bf 9}, 1753 (1992); 
    D. H. Bernstein and K. P. Tod, {\em Phys. Rev.}  {\bf D 49}, 2808 (1994).
\bibitem{VirKN} K. S. Virbhadra, {\em Phys. Rev.} {\bf D42}, 1066 (1990); 
                                  {\em Phys. Rev.}  {\bf D42}, 2919  (1990).
\bibitem{CoopRi} F. I. Cooperstock and S. A. Richardson, in {\em \ Proc.
        4th Canadian Conf. on General Relativity and Relativistic Astrophysics}
        (World Scientific, Singapore, 1991).
\bibitem{VirOthFlat}
  K. S. Virbhadra, {\em Phys. Lett.} {\bf A157}, 195 (1991);
  K. S. Virbhadra, {\em Mathematics Today} {\bf 9}, 39 (1991); 
  K. S. Virbhadra, {\em Pramana - J. Phys}. {\bf 38}, 31 (1992); 
  K. S. Virbhadra and J. C. Parikh, {\em Phys. Lett.} {\bf B317}, 312 (1993); 
                                    {\em Phys. Lett.} {\bf B331}, 302 (1994); 
  K. S. Virbhadra, {\em Pramana - J. Phys}. {\bf 44}, 317 (1995); 
 A. Chamorro and K. S. Virbhadra, {\em Pramana - J. Phys.} {\bf 45}, 181 (1995);
 A. Chamorro and K. S. Virbhadra, {\em Int. J. Mod. Phys.} {\bf D5}, 251 (1996);
 K. S. Virbhadra, {\em Int. J. Mod. Phys.} {\bf A12}, 4831 (1997); 
 S. S. Xulu, {\em Int. J. Theor. Phys.} {\bf 37}, 1773 (1998); 
 S. S. Xulu, {\em Int. J. Mod. Phys.} {\bf D7}, 773 (1998).
\bibitem{VirOthNonFlat}
  N. Rosen and K. S. Virbhadra, {\em Gen. Relativ.  Gravit.}, {\bf 25}, 
          429 (1993); 
  K. S. Virbhadra, {\em Pramana - J. Phys}. {\bf 45}, 215 (1995);
  N. Rosen, {\em Gen. Relativ. Gravit.}, {\bf 26}, 319 (1994); 
  F. I. Cooperstock, {\em Gen. Relativ. Gravit.}, {\bf 26}, 323 (1994); 
 F. I. Cooperstock and M. Israelit, {\em Found. of Phys.} {\bf 25}, 631 (1995);
 N. Banerjee and S. Sen, {\em Pramana-J. Physics}, {\bf 49}, 609 (1997);
 S. S. Xulu, {\em Int. J. Theor. Phys} (in press) gr-qc/9910015;
 S. S. Xulu, {\em Int. J. Mod. Phys.} {\bf A} (in press), gr-qc/9902022.

\bibitem{Changetal}  C. C. Chang, J. M. Nester and C. Chen, {\em Phys. Rev.
           Lett.} {\bf 83}, 1897 (1999).

\bibitem{CodeFe}  J. M. Cohen, \ F. de Felice, {\em J. Math. Phys.},{\bf 25},
    992 (1984).
\bibitem{Lessner}  G. Lessner, {\em Gen. Relativ. Gravit}. {\bf 28}, 527
       (1996).
\bibitem{CoopHypoth} F. I. Cooperstock, {\em Mod. Phys. Lett.} {\bf A14} 
             1531 (1999).
\bibitem{Tod} K. P. Tod, {\em Proc. Roy. Soc. London}  {\bf A388}, 467
   (1983).
\end{references}
\end{document}